\documentclass[%
preprint,
 amsmath,amssymb,
prb,
]{revtex4-1}

\usepackage{graphicx}
\usepackage{dcolumn}
\usepackage{bm}
\usepackage{xcolor}
\usepackage{ulem}

\usepackage{natbib}

\begin{document}


\title{Stacking and interlayer electron transport in MoS$_2$}

\author{Teresa Cusati}
\email{teresa.cusati@for.unipi.it}
\affiliation{Dipartimento di Ingegneria dell'Informazione, Universit\`a di Pisa, Via G. Caruso 16, 56122 Pisa, Italy\\}

\author{Alessandro Fortunelli}
\affiliation{CNR-ICCOM, Consiglio Nazionale delle Ricerche, Via Moruzzi 1, 56124 Pisa, Italy\\
}%

\author{Gianluca Fiori}
\affiliation{Dipartimento di Ingegneria dell'Informazione, Universit\`a di Pisa, Via G. Caruso 16, 56122 Pisa, Italy\\
}%

\author{Giuseppe Iannaccone}
\affiliation{Dipartimento di Ingegneria dell'Informazione, Universit\`a di Pisa, Via G. Caruso 16, 56122 Pisa, Italy\\
}%

\date{\today}

\begin{abstract}
In this work, we investigate the effect of the stacking sequence in MoS$_2$ multilayer systems on their electron transport properties, through first-principles simulations of structural and electron transport properties. We show that interlayer electron transport is highly sensitive to the stacking sequence of the multilayers, with specific sequences producing much higher electron transmission due to larger orbital interactions and band structure effects. These results explain contrasting experimental evidence on interlayer transport measurements as due to imperfect structural control, provide insight on modeling and suggest ways to improve the performance of electron devices based on MoS$_2$ multilayer systems via multilayer structure engineering.
\end{abstract}

\maketitle


\section{\label{sec:intro}Introduction}

Transition metal dichalcogenides (TMDs) and, in particular, layered TMDs, have received great attention in recent years due to their appealing physical and electrical properties, enabling their use in transistors, photodetectors and other electron devices. \cite{Kis2011, Wang2012, Georgiou2013, Kis2011B, Padilha2014} Assuming defect-less individual layers, the key structural degree of freedom defining these materials is the relative arrangement (stacking) of the layers. The relative weakness of van der Waals forces, responsible of cohesion, allows sliding or rotation of adjacent layers, resulting in different stacking sequences. Bulk MoS$_2$ crystallizes in two different polytypisms, 2H and 3R, differing in the stacking orientation of the layers. \cite{He2014} The electronic properties of these materials are rather sensitive to morphology and interlayer interaction, \cite{Mak2010, Song2015, Lu2012, Kumar2013} with band gap energy and direct-to-indirect band gap transition depending on the number of layers and interlayer distance. \cite{Zande2014, Han2011,Fang2014, Yeh2015, Sharma2014} In the literature, several experimental and theoretical studies have thus been performed, \cite{He2014, Yan2015, Yeh2016, Wang2015, Huang2014, Fan2016,Levita2014} to investigate these phenomena. However, most investigations analyzed electronic properties (band gap variation, interlayer coupling and spectroscopic responses) as a function of the rotation angle of MoS$_2$ bilayer and trilayer, \cite{Zhou2017, Zande2014, Tan2016, Cao2015, He2014, Yan2015} whereas few experimental and theoretical works have dealt with the effects of rotation angle on interlayer resistance. \cite{Tan2016,Zhou2017,Sengupta2016} Therefore, there are still important issues to be overcome at experimental level for rotated/slided MoS$_2$, i.e., how to control the stacking orientation and the consequent formation of Moir\'e patterns and the presence of stacking faults, and how these features affect quantities related to electric transport such as Schottky barrier and the transmission coefficient. It is thus still so far unclear which is the most favorable stacking for electron transport in between the layers and which is the range of variation. A very recent experimental/theoretical work \cite{Shinde2018}  has made important progress in this sense by achieving better control of stacking in multilayered MoS$_2$ and how this affect the optical properties of these systems, focusing on 2H and 3R stacking and a combination of both. In such a context, theory can be then very important to guide the experiments to configurations otherwise neglected, and which may exhibit different transport features. The ideal goal is then to develop a simple model to describe the physics of electron transport under rotation/sliding, thus providing clues about the factors to be considered in the device design. 

To shed light on this topic, in this work we have performed a systematic first-principles investigation of interlayer electron transport in multilayer MoS$_2$ structures. We consider all possible high-symmetry stacking sequences, in which sulfur atoms of one layer occupy hollow or on-top sites of the neighboring layer, which represent limiting cases of arrangements produced from the rotation/sliding of one MoS$_2$ layer with respect to the other. We find a maximum of transmission efficiency for `on top' and one specific `hollow' configuration (not corresponding to bulk stacking), which we explain on the basis of orbital interactions between layers and consequent changes in the band structure. These results provide a solid ground for the analysis of vertical transport features in MoS$_2$ and other TMD materials, and their applications to electron devices. 

Several studies (at both experimental and theoretical level) have considered arbitrarily aligned bilayer MoS$_2$ systems and investigated the effect of interlayer stacking on the electronic properties. \cite{Zande2014,Yeh2016,Liu2014} Theory in particular \cite{He2014, Cao2015, Zhou2017, Zande2014, Levita2014} has focused on the dependence of interlayer coupling on the twist angle, finding that such coupling reaches maximum values for the 2H and 3R phases and decreases for intermediate angles. This finding has been explained in terms of the change of interlayer separation with the twist angle, which determines the overlap among Molybdenum and Sulfur orbitals of different layers and thus the band splitting around the $\Gamma$ point. Vertical electron transport has instead been considered in only few works. In particular, in a recent theoretical work by Zhou et al., \cite{Zhou2017} they have found that interlayer misorientation of bilayer MoS$_2$ suppresses vertical electron transport and consequently leads to an increase in electron resistivity with the twist angle. \cite{Zhou2017} As we will show in the following, by using highly symmetric configurations we can here explore sliding/rotation angles of multilayered MoS2 not considered in this previous work, and show that for specific stacking configurations the opposite effect of an {\em increase in electric transport} can occur. Significant effects of different stacking orders on carrier effective masses and carrier transmission were also observed in other layered systems such as black phosphorous.\cite{Sengupta2016} One problem to be faced in computational studies is that mis-oriented layers become incommensurate and in principle require the use of big supercells in the simulation and therefore have high computational cost. To overcome this issue, in the present work we adopt a model that considers only limiting stacking cases, thus reducing the computational effort to a manageable level while still providing rigorous clues about the physics behind electron transport in such systems.

\section{\label{sec:model}Model}

\begin{figure}[ht!]
\begin{center}
\includegraphics[scale=0.6,clip]{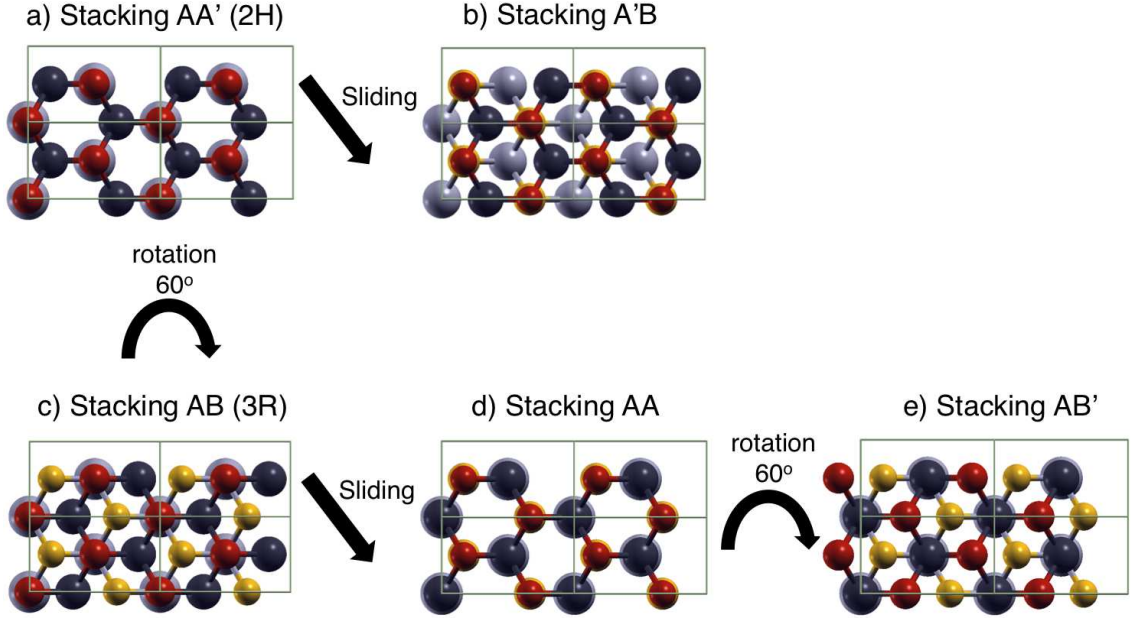}
\caption{\label{fig:1} Top view of the schematic atomistic representation of stacking configurations of MoS$_2$ bilayers and the conversion mechanism among them: a) AA' eclipsed stacking; b) A'B staggered stacking; c) AB staggered stacking; d) AA eclipsed stacking; e) AB' staggered stacking. The orthorhombic unit cell is replicated and the atomic color was set different in every layer, for visualization reasons: yellow and light gray for S and Mo atoms of the lower layer, respectively, and red and dark gray for S and Mo atoms of the upper layer, respectively. A lateral view of the same stacking configurations is shown in the Figure S1 of Supplemental Material.\cite{SuppMat}}
\end{center}
\end{figure}

Our simulations employ unary cell systems replicated in 3D, as illustrated in Figure \ref{fig:1}. These stacking scenarios exhaustively sample all high-symmetry configurations in which sulfur atoms of one layer occupy hollow or on-top sites of the neighboring layer, thus representing limiting cases of arrangements produced from the rotation/sliding of one MoS$_2$ layer with respect to the other. The choice of unary cell configurations allows us to perform rigorous transport calculations with a reasonable computational effort. For the nomenclature of the systems under study we follow the scheme used in ref \cite{He2014} and illustrated in Figure \ref{fig:1}:

\renewcommand{\theenumi}{\alph{enumi}}
\begin{enumerate}
    \item{AA' eclipsed stacking, with Mo over S;}
    \item{A'B staggered stacking, with S over S and Mo in hollow position with respect to the layer underneath, obtained from a diagonal sliding of one layer in the AA' stacking;}
    \item{AB staggered stacking, with one of the S over Mo and the other one in hollow sites, resulting of the rotation of 60 degrees of one layer in the AA' stacking;}
    \item{AA eclipsed stacking with Mo over Mo and S over S, as a result of the diagonal sliding of one layer in the stacking AB;}
    \item{AB' staggered stacking, with Mo over Mo and S lying on hollow sites of the underneath layer, obtained from the rotation of 60 degrees of one layer in the AA stacking.}
\end{enumerate}

It should be noted that AA' corresponds to the stacking in the bulk 2H phase, whereas AB approximately corresponds to the stacking in the bulk 3R phase (our replicated system does not exactly correspond to the 3R phase whose unit cell is a trilayer in which the top and bottom are staggered, not eclipsed). 
 
These stacking sequences can be classified in two groups, taking into account the epitaxy of the sulfur atoms between the two layers of MoS$_2$:

\renewcommand{\theenumi}{\roman{enumi}}
\begin{enumerate}
    \item{`hollow' (staggered) configurations, where S atoms of one layer are placed in hollow sites of the layer underneath or on top of Mo atoms. These configurations utilize the experimental interlayer distance of 2.98 {\AA} as in the bulk MoS$_2$. \cite{He2014, Bronsema1986} In this category we include AA', AB and AB' stacking (see Figure 1 (a), (c) and (e) respectively). \cite{Noteconf}}
    \item{`on top' (eclipsed) configuration, where S atoms of both layers are located in on-top sites, with an interlayer distance of 3.50 {\AA}, corresponding to the experimental equilibrium distance between S atoms in different layers. We consider the choice of this distance a good tradeoff to describe this configuration (see Supplemental Material for details. \cite{SuppMat}) In this group we include stacking sequences A'B and AA  (see Figure \ref{fig:1} (b) and (d) respectively).} 
\end{enumerate}

For these configurations, we have performed first-principles DFT simulations with the Quantum Espresso package, \cite{QE2009} using a plane wave basis set, a gradient-corrected exchange-correlation functional (Perdew-Burke-Ernzerhof (PBE)), \cite{Perdew1996} scalar-relativistic ultrasoft pseudopotentials (US-PPs) \cite{Vanderbilt1990} and including Grimme's DFT-D2 dispersion correction. \cite{Cheng2011,Grimme2006,Barone2008} Spin-orbit coupling has not been taken into account, because of its small effect on the band structure. \cite{Zhou2017,Tan2016} For more details about calculations, see Supplemental Material.\cite{SuppMat}

\section{\label{sec:results}Results}

Table {\ref{tab:table1}} reports the electronic properties (valence band maximum (VBM), conduction band minimum (CBM), direct and indirect band gaps and relative energy with respect to the AA' stacking) of extended bilayer MoS$_2$ systems for the different stacking configurations. VBM shows a marked variation between `hollow' configurations and `on top' ones, reaching the highest values for the AA and A'B stacking orders. These results are in agreement with previous works \cite{Han2011,He2014}, thus validating our approach. Note that in the `on-top' configurations the S atoms are eclipsed and the interaction between their $p_z$ orbitals is maximum (this will be important in the following).  Table {\ref{tab:table1}} also includes the interlayer coupling quantities for the CBM at $\hbox{\rm K}_C$ point and VBM at $\hbox{\rm K}_V$ and $\Gamma_V$ points. The results for the AA' (2H phase) are in agreement with the values reported in ref \cite{Zhou2017}. Here, what is to be noted is that AB' stacking presents the highest interlayer coupling for the VBM at $\Gamma_V$ and the lowest $\hbox{\rm K}_V$ hinting at the specific difference of this phase as discussed below. \cite{Cao2015}

\begin{table}[h!]
\begin{center}
\caption{\label{tab:table1} Electronic properties of extended bilayer MoS$_2$ systems at different stacking order: conduction band minimum (CBM), valence band maximum (VBM), direct and indirect band gaps. Relative energy is obtained with respect to th AA' stacking. Interlayer coupling (in eV) of the CBM at $\hbox{\rm K}$ point and the VBM at $\hbox{\rm K}$ and $\Gamma$ points are also reported.}
\vspace{0.5cm}
\scalebox{0.90}{
\begin{tabular}{|c|c|c|c|c|c|c|c|c|c|}
\hline
       & Stacking & CBM (eV) & VBM (eV) & $\Delta E$ ($\hbox{\rm K}$-$\hbox{\rm K}$) & $\Delta E$ ($\hbox{\rm K}$-$\Gamma$) & Relative & $\hbox{\rm K}_C$  & $\hbox{\rm K}_V$  & $\Gamma_V$  \\
       &          &          &          & Direct gap (eV) & Indirect gap (eV)   & energy (eV)  &    (eV)     &    (eV) &        (eV)       \\
\hline
\hline
       &    AA'   &  4.31    &   5.41   &     1.61    &       1.10      & 0.000 &   0.000  & 0.087 & 0.668   \\
Hollow &    AB    &  4.33    &   5.38   &     1.57    &       1.05      & -0.001 &         0.066 & 0.089 & 0.686   \\
       &    AB'   &  4.32    &   5.39   &     1.64    &       1.07      & 0.014 & 0.009  & 0.000  & 0.701   \\
\hline       
  Top  &    AA    &  4.33    &   5.60   &     1.60    &       1.27      & 0.060 &        0.004  & 0.038 & 0.475   \\
       &    A'B   &  4.31    &   5.60   &     1.64    &       1.29      & 0.055  & 0.001  & 0.001  & 0.477   \\       
\hline
\end{tabular}
}
\end{center}
\end{table}

From Figure \ref{fig:3} and Figure S2 of the Supplemental Material \cite{SuppMat} in tune with the discussion in Ref. \cite{Zhou2017}, the predominant contribution to the conduction band at $\hbox{\rm K}$ is due to $d_{z^2}$ orbital of Mo atoms, while the valence band is principally made of $d_{x^2-y^2}$ and $d_{xy}$ orbitals at $\hbox{\rm K}_V$ and $p_z$ orbitals of S atoms and $d_{z^2}$ orbitals of Mo atoms at $\Gamma_V$. The interlayer coupling at $\hbox{\rm K}_C$ for AA' stacking is then an order of magnitude smaller than at $\hbox{\rm K}_V$. Moreover, in the configurations analyzed in Ref. \cite{Zhou2017}, in the rotated geometries there is a destructive interference in the phases of the wave function, leading to a slow decrease of the coupling $\hbox{\rm K}_C$. In our case, this is not observed because no symmetry breaking occurs, $\hbox{\rm K}_C$ has small values for all stackings except for the AB one, where the contribution of the $d$ orbitals to the conduction band is lower. This is then reflected in the larger values of $\hbox{\rm K}_C$.

Building on the SCF calculations of the electronic structure of the configurations depicted in Figures S3 of the Supplemental Material \cite{SuppMat}, electron transmission simulations were performed, using the PWCOND module \cite{Smogunov2004,Choi1999} included in the Quantum Espresso package \cite{QE2009} (details about these simulations are provided in the Supplemental Material \cite{SuppMat}). 

Figure \ref{fig:2} shows the comparison among the vertical transmission coefficients of the structures described above. The most relevant insights that can be drawn from an inspection of this Figure are:

\begin{itemize}
    \item{AB and AA' stacking configurations show lower transmission coefficient below the Fermi level,}
    \item{electron transmission is higher for the structures where the sulfur atoms are sited in `on top' position (stacking AA and A'B), but also for the hollow structure AB',}
    \item{the transmission curves above the Fermi level are similar to each other except for the AB stacking, which again presents a lower conductance.}
\end{itemize}

\begin{figure}[ht!]
\begin{center}
\includegraphics[scale=0.57,clip]{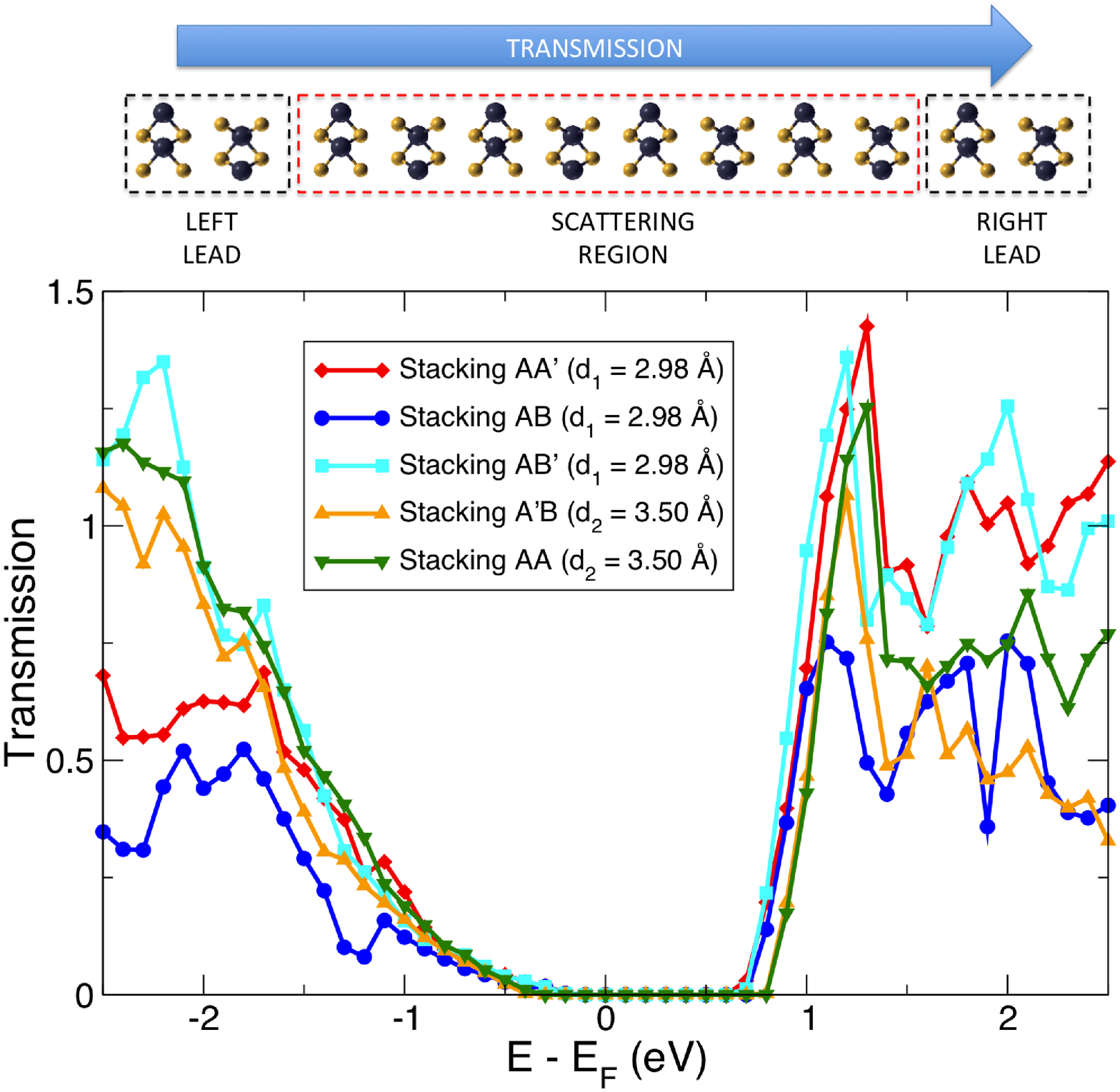}
\caption{\label{fig:2} Transmission coefficient as a function of energy computed for the contacts showed in Figure \ref{fig:1}.}
\end{center}
\end{figure}

\begin{figure}[ht!]
\begin{center}
\includegraphics[scale=0.47,clip]{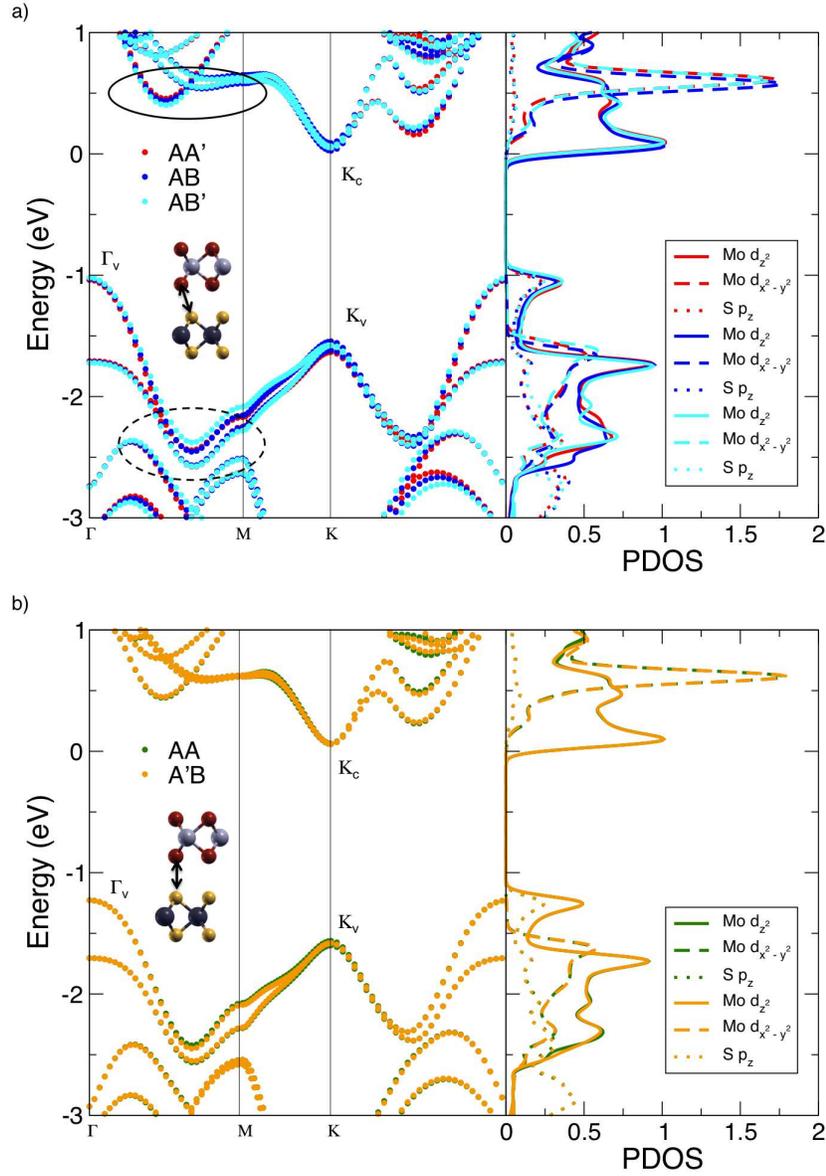}
\caption{\label{fig:3} Band structure and PDOS of selected orbitals ($d_{z^2}$ and $d_{x^2-y^2}$ for Mo atoms and $p_z$ for S atoms) of a) `hollow and b) `on top' stacking in the energy range close to the Fermi energy, for a hexagonal cell. Coupling parameters at $\Gamma$ and $\hbox{\rm K}$ points are indicated inside each picture. In (a), black circles point out differences in the topmost valence band and bottommost conduction band: highest valence band at $\Gamma$-$\hbox{\rm M}$ region is indicated with a dashed circle, lowest conduction band in $\hbox{\rm K}$-$\Gamma$ region is indicated with a full line circle.}
\end{center}
\end{figure}

\begin{figure}[ht!]
\begin{center}
\includegraphics[scale=0.5,clip]{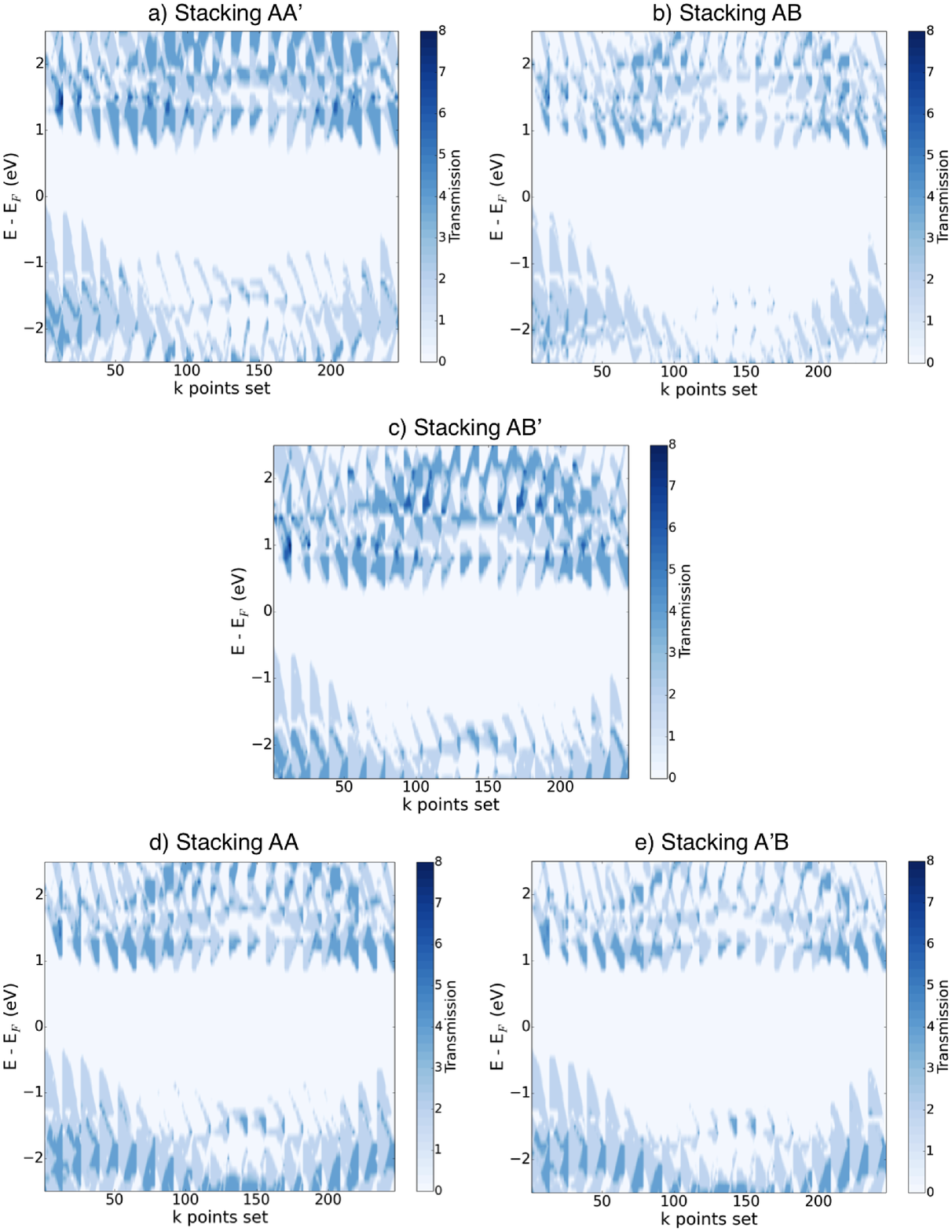}
\caption{\label{fig:4} Color plot of transmission coefficient as a function of energy and $k$-points sets defined from the uniform k-mesh of the orthorhombic cell in the electron transport simulations with PWCOND module, for `hollow' configurations: (a) AA', (b) AB and (c) AB' and `on top' configurations: (d) AA and (e) A'B.}
\end{center}
\end{figure}

These results can be explained as follows. The `on-top' configurations achieve a larger conductance with respect to `hollow' configurations, because of a better inter-layer overlap among $p_z$ orbitals of the S atoms in close contact. Among `hollow' configurations, the behavior is different in the conduction and valence bands. In the valence band the AB' stacking reaches the largest transmission, appreciably higher than the AA' (2H) and AB (3R) phases especially at 1.5 eV below the Fermi energy. An inspection of the band structure and the contribution of some orbitals (in particular $d_{z^2}$ and $d_{x^2 - y^2}$ of the Molybdenum and $p_{z}$ of Sulfur atoms) to the PDOS in Figure \ref{fig:3} shows that in this energy region and close to the $M$ point ($\Gamma$-$\hbox{\rm M}$ region) of the Brillouin Zone, the energy difference between the topmost valence band and the bottommost conduction band is clearly larger for the AA' and AB stacking, with respect to AB'. This difference occurs also in other regions of the Brillouin Zone (for instance, in the k-path of the orthorhombic cell), although it is not uniform, and – together with the important role played by the Coulomb interactions between the layers – is in part responsible for the greater energetic stability of the AA' (2H) and AB (3R) phases (which are indeed those observed as bulk crystals) but also reduces conductance because of a reduced overlap among electronic wave functions. Significantly, in the conduction band above the Fermi energy the AB', AA and A'B stacking presents a larger transmission with respect to the bulk phases, but at these energies there is a compensation in the $\Gamma$-$\hbox{\rm K}$ region of the Brillouin Zone in which the AA' phase exhibits a smaller band difference/larger transmission, so that the overall increase in transmission is marginal. The reasons of such behavior \cite{Zhou2017} lie in a `charge compression' effect. In the configurations with an interlayer distance of 2.98 {\AA}, the repulsion among the electronic clouds of the S atoms pushes the valence band at higher energy, especially at the $\Gamma$ point, producing a decrease of the indirect band gap, \cite{Howell2015, Padilha2014, Galli2007} but also of the transmission coefficient. Instead, shifting the stacking relationship to the `on-top' configuration also optimizes orbital overlap, thus leading to nearly ideal transmission. Rotating/sliding and decoupling the MoS$_2$ layers decreases charge compression effects, and is thus singled out as an efficient tool to significantly improve transport properties. Additionally, a similar effect can also be obtained in the AB' configuration at an interlayer distance of 2.98 {\AA}, in which the energy gap around to the M point of the Brillouin Zone (encircled regions in Figure \ref{fig:3}) is reduced due to interference effects and transmission consequently improved.

To confirm the previous analysis, we have also decomposed the transmission in terms of contributions of individual $k$-points set to the total transmission coefficient, as a function of the energy. The k-points set corresponds in this case to the uniform mesh of k-points in an orthorhombic cell used in the transport simulations. Every set is composed by two coordinates: $k_x$ and $k_y$, perpendicular to the propagation direction $z$. In this way it is possible to evaluate transmission in a particular region of the k-space. Accordingly to the color plots depicted in Figure \ref{fig:4}, there is a lower contribution of specific $k$-points in proximity of the $\hbox{\rm K}$ point of the Brillouin Zone (BZ) for the AB stacking configuration with respect to the AA' and AB' ones, in tune with a reduced transmission. 

\begin{figure}[ht!]
\begin{center}
\includegraphics[scale=0.6,clip]{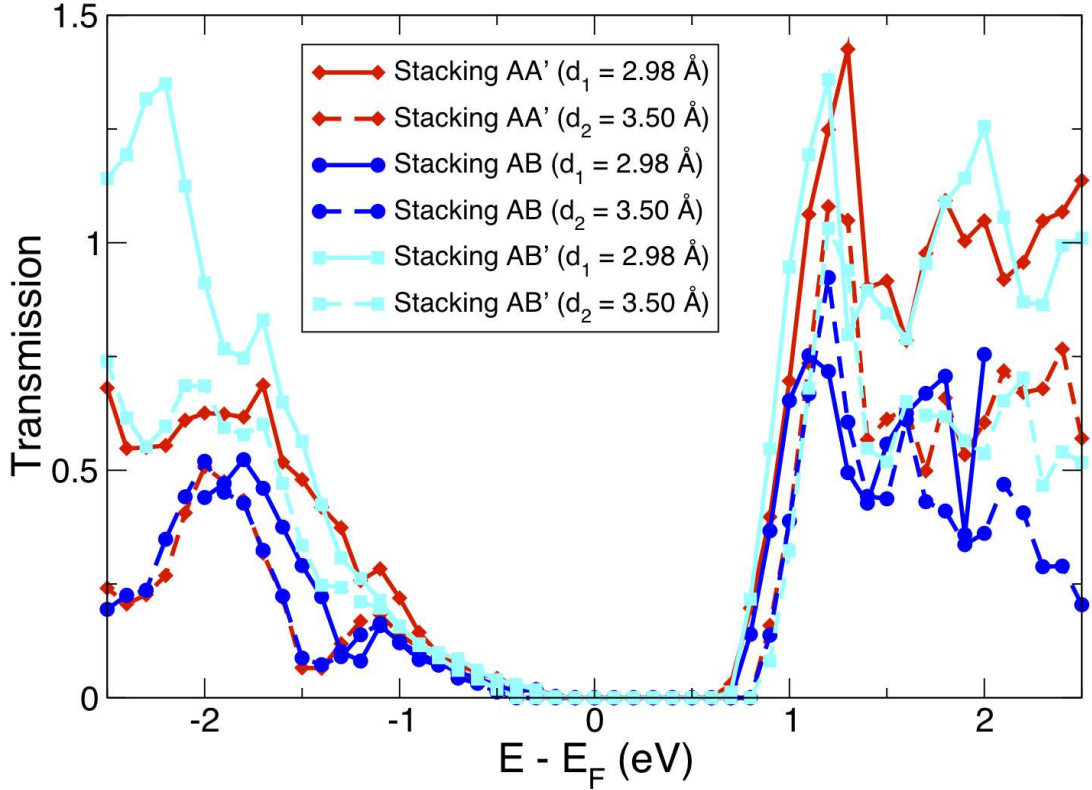}
\caption{\label{fig:5} Comparison of the transmission coefficient as a function of energy computed for the epitaxial contacts showed in (a), (b), (c) of Figure S3, considering an interlayer distance of d$_1$ = 2.98 \AA (full lines) and d$_2$ = 3.50 \AA (dashed lines).}
\end{center}
\end{figure}

To complete our analysis, we finally calculated transmission coefficients for the `hollow' configurations but considering as interlayer distance both 2.98 and 3.50 {\AA}, and show the corresponding results in Figure \ref{fig:5}. The interlayer distance depends on the interlayer rotation in a non-uniform way. In any case, although Moir\'e patterns can form, regions with on-top or near on-top configurations cannot in general be avoided. Our model simulations thus mimic the situation in which an incoherent stacking among MoS$_2$ layers create simultaneously regions with `hollow' and `on top' stackings (and intermediate arrangements), all at the interlayer distance of 3.50 {\AA} enforced by the presence of the `on-top' regions and the associated S-S interlayer repulsions. In general, transmission decreases when increasing interlayer distance, because of the decrease in overalp matrix elements among layers. 

\section{\label{sec:conclu}Final remarks}

In summary, we have studied the effects of the stacking order on the electronic structure and electron transport features in multilayer MoS$_2$ via first-principles simulations. As in previous work, the electronic structure of bilayer MoS$_2$ shows significant differences in the indirect band gap and interlayer coupling parameter, depending on the stacking sequence considered. Here, we additionally show that this appreciably affects transport properties. For `on top' stacking, in which the overlap interaction between the sulfur atoms is the highest, we find much larger electron transmission with respect to bulk-like `hollow' stacking configurations. Moreover, also within the `hollow' stacking configurations, there exists one, corresponding to AB' stacking, which presents larger transmission especially below the Fermi level with respect to AA' and AB stackings (corresponding to 2H and 3R bulk phases). This behavior is explained as due to a decrease of charge compression and a tuning of orbital interactions between sulfur atoms belonging to different MoS$_2$ layers.

We conclude that van der Waals stacking in 2D layered materials entails a degree of freedom in the epitaxial relationships among stacked monolayers that affects not only their electronic structure but also their transport behavior -- this latter dependence is more subtle than the former due to the larger sensitivity of transport to off-diagonal elements of the density matrix and details of the band structure. Theory provides an important insight into this phenomenon, and can guide experiment to single out unexpected effects. In particular, we find a larger value of electron transmission for the `on top' stacking in which however the interlayer distance is larger with respect to the `hollow' ones, and we also find a near optimal transmission in one specific `hollow' configurations which present a significant orbital overlap but reduced band splitting at the $\Gamma$-$\hbox{\rm K}$ points. Interestingly, we thus find an increase of the electron transmission for properly engineered $misoriented$ configurations. These results provide additional insights with respect to previous work, including configurations not considered before (such as 'on top' configurations or AB stacking) and extend our knowledge on the dependence of transport properties of MoS$_2$ multilayer systems upon rotation/sliding, thus suggesting the importance of MoS$_2$ stacking to achieve control of conduction and consequently performance of MoS$_2$-based electronic devices.

This work was supported by EC H2020 program through the Graphene Flagship Core 2 (Contract 785219).

\bibliographystyle{apsrev4-1}
\bibliography{Biblio.bib}

\end{document}